# Growth of toric domains in the smectic phase of oxadiazoles.

A. Sparavigna[1], A. Mello[1] and B. Montrucchio[2]

[1] Dipartimento di Fisica, Politecnico di Torino

[2] Dipartimento di Automatica ed Informatica, Politecnico di Torino

C.so Duca degli Abruzzi 24, Torino, Italy

Keywords: Liquid crystals, phase transitions

**Abstract**

In a previous paper (Phase Transitions, 80(9), 987, 2007) we discussed the smectic and nematic textures of some mesomorphic oxadiazole compounds with terminal Cl-substituent. Optical microscope investigations showed a very interesting behaviour of smectic and nematic phases; the smectic phase has fan-shaped and toric textures and the nematic phase has spherulitic domains, which disappear as the sample is further heated, the texture changing into a smooth one. Here, we investigate four oxadiazole compounds with the same structure but terminal Br-substituent. The behaviour of the smectic and nematic phases is like that observed in the compounds with Cl. The focus of the paper is on the growth of toric domains from the nematic melt and on the role of defects in the domain structure.

**Introduction**

In a previous paper [1], hereafter referred as I, we investigated a family of mesomorphic oxadiazoles with asymmetrical molecular structures. The family consisted of oxadiazoles [2-4], containing terminal Cl-group and phenyl-cyclohexane fragment together with biphenyl analogues. The temperature range of the mesophases and the existence of a smectic phase are conditioned not only by the chemical structure of the substituents, but also by their position with respect to the oxadiazolic ring. The microscope observations, discussed in I, showed interesting smectic and nematic phases of the oxadiazole compounds. In the smectic phase, fan-shaped and toric textures have been observed, sometimes with periodic instability in the toric domains. Moreover, it was also remarkable the behaviour of the nematic phase in the compounds possessing both smectic and nematic mesophases; the nematic exhibits a texture transition, that is a change driven by the temperature, of the texture observed by means of the polarised light microscope. Below the temperature of the texture transition, the nematic shows spherulitic domains, similar to those observed in twisted nematics close to the smectic A phase transition [5]. Above the transition temperature, the nematic assumes the usual schlieren appearance. Texture transitions have been previously observed inside the nematic range of some alkyloxybenzoic and cyclohexane acids [6-13]. The common explanation for the presence of a transition in the nematic range of alkyloxybenzoic acids is based on the existence of cybotactic clusters favouring a local smectic order in the nematic melt [14,15]. In I, we suggested the existence of cybotactic clusters in the nematic phase of oxadiazole compounds, clusters providing the local smectic order responsible for the texture transition.

Here we will discuss other members of the oxadiazole family, with terminal Br atom substituting the Cl atom. The focus of this paper will be on the formation of toric domains in the smectic phase from the nematic melt. Because the transition on cooling from nematic in smectic phase can be slowly driven, a detailed record of the growth of the toric domains is possible. A clear correlation between the position of the domain walls in the nematic and of the toric structures in the smectic is easily recognisable. First of all, let us give a short description of the compounds.

**Oxadiazole compounds with terminal Br substituent.**

The members of the oxadiazole family we are discussing in this paper are those with structures reported in Fig.1. We investigate four samples with Br attached at a terminal phenyl ring. As in I, the oxadiazole compounds, prepared by the liquid crystal research group at the Organic Intermediates and Dyes Institute of Moscow, have an asymmetric heterocycle structure. Table 1 reports the transition temperatures on heating and cooling of these compounds. A noticeable influence of the molecular structure is evident from data reported in Table 1: the existence of the smectic phase is influenced by the position of Br atom on the phenyl ring. The same influence was observed in the compounds discussed in I. The length of the alkyl tails is playing a slight role in changing the transition temperatures.

All the samples were inserted in the cell when the material was in the isotropic phase. The walls of liquid crystal cells are untreated clean glass surfaces. No treaments were done to favour planar or homeotropic alignments. Spacers are not used between cell walls and then the cell thickness is few microns. The liquid crystals are heated and cooled in a thermostage and textures observed with a polarised light microscope. Let us start the discussion from compound $S_1$ of Figure 1.

**Compound $S_1$**

This sample has a nematic and a smectic phase. The temperature transitions are reported in the Table 1. The smectic phase can have different texture arrangements in the cell: in some areas of the cell the smectic assumes a homeotropic configuration, in other regions the texture is composed of fan-shaped or toric domains. The observed toric domains have a large size, approximately 0.05 mm (see Fig.2). The smectic fan domains have the typical features of those proposed by the images of the Demus and Richter collection [16] for the smectic A fan textures. As in I, we assume then a smectic A phase for these oxadiazoles.

In the nematic phase, in the temperature range below 150°C, on heating and on cooling, we observe the presence of domains with bent contours mixed with quasi-homeotropic domains. If the temperature is higher than 150°C, the homeotropic regions disappear and a planar texture grows, occupying the totality of the cell. The low temperature nematic texture can be considered as a sort of spherulitic nematic texture. We mean by spherulitic nematic texture, the particular arrangement of domains with bent contours that we observed in the oxadiazole compounds C and D discussed in I. The spherulitic texture is observed in sample $S_2$ too and then we postpone the detailed discussion in the next subsection.

A sequence of textures obtained on cooling compound $S_1$ from the isotropic melt is shown in Fig.3: a planar texture at high temperature, the spherulitic nematic where homeotropic regions arise, and the toric domains in smectic texture. The smectic phase has a notable hysteresis on cooling, because it is observed till down 85°C. Under this temperature, the sample becomes a crystal.

**Compound $S_2$**

The sample has a nematic and a smectic phase. The temperature transitions are reported in Table 1. As for compound $S_1$, we observe a smectic phase with different textures: homeotropic with clusters of toric domains and fan-shaped domains. In this case too, we suppose a smectic A phase. A noticeable hysteresis of the smectic phase is displayed by the sample on cooling, because smectic is observed till down 80°C. In the temperature range of the nematic phase, below 175°C, on heating and on cooling, we observe a nematic with spherulitic domains. It is not possible to find a well-defined texture transition temperature in the nematic range, but between 175 and 180°C the texture turns into a planar one. To have a good recording of the transition, we chose a region with a small spherulitic domain in the homeotropic black: the Fig.4 shows the texture change as temperature increases. Fig.5 is instead representing how the nematic phase with spherulitic domains transforms itself into a toric smectic texture. The transition is rather slow, occupying approximately one degree.

$S_1$ and $S_2$ behave like compounds C and D, as shown in I. In that paper we discussed the nematic phase with texture transitions inside, from the low temperature nematic sub-phase with spherulitic arrangement of domains, into the high temperature sub-phase where domains disappear and the texture becomes smooth. The spherulitic texture is due to a splay-bend configuration of the nematic director, giving domains with bent contours. As previously told, a reasonable explanation for the existence of a texture transition is the presence in nematic melts of cybotactic clusters with local smectic order. Let us imagine these clusters of molecules as seeds of smectic order in the nematic fluid. In the previous paper we invoked the presence of such clusters also in the oxadiazole compounds displaying the texture transitions. The local smectic order would be responsible for a spherulitic nematic texture: this means that the nematic phase exhibits different local orders at low and high temperatures and that the different orders are revealed by the unlike textures observed with optical microscopy. When temperature is low enough, the material can achieve the smectic order with a texture full of toric domains.

The behaviour of the nematic phase of alkyloxybenzoic acids is more or less the same: the low temperature nematic texture is composed of very small domains and looks like a granular pattern and the smectic phase has a granular structure too. A strong paramorphosis is existing between the low temperature nematic phase and the smectic phase [6,7,10]. Probably, it is the role of the anchoring at the cell walls of cybotactic clusters, which is relevant for the paramorphosis. We will deal with the paramorphosis of oxadiazole compounds in a following section, in which we shall discuss the growth of the toric domains.

**Compounds $S_3$ and $S_4$**

These two compounds have only the nematic mesophase. In the nematic range there is no evidence of a texture transition. This is the same behaviour of compounds E,F and G studied in I. Because texture transitions are not observed, we suppose that cybotactic clusters cannot grow in the nematic melt as temperature decreases, and this is consistent with the fact that the medium does not achieve the order of a smectic phase. As for compounds with Cl-substituent, the growth of local smectic order can be hindered by a frustration in the molecular aggregation that does not allow the arrangement into a layered structure of molecules. The responsible can be the position of the Br (or Cl) atom (see Fig.1).

**Growth of toric domains.**

A sort of paramorphosis is existing between the low temperature nematic phase and the toric smectic of the observed oxadiazole compounds. In paramorphosis, it is the anchoring of molecules at the cell walls, which plays the main role. In this section, we discuss how toric domains are growing from the domains of spherulitic nematic. The transition from nematic into smectic for both compounds, $S_1$ and $S_2$, can be so slowly driven, that the feature changes can be recorded without special devices.

The toric domain is a focal conic domain where the ellipse is degenerated in a circle and the hyperbola is degenerated in a straight-line [17]. The upper part of Fig.6 is a detail of the toric texture of $S_2$: in the lower part, the director field in a vertical cross-section of the cell is drawn. The structure of the smectic layers is also drawn. The circle of the focal domain is parallel to the cell wall and it is represented in the vertical cross-section by the red line; the blue line perpendicular to the cell walls is representing the degenerated hyperbola.

Let us imagine a small part of a thin liquid crystal cell in the nematic phase, with a defect in its structure. As in Fig.7 on the left, the defect can be the boundary between planar domains. Lowering the temperature, the material phase changes into the smectic phase; the material must arrange itself in smectic planes. Of course, molecules are anchored at the cell walls and the anchoring is preserving the local orientation of director. From the sequence of Fig.7, it seems that the arrangement of the molecules is maintained in the passage from nematic in smectic at: (a) the boundaries between the homeotropic region and a quasi-planar area, and (b) between two quasi-planar domains. In the domains then, the molecules reorient to achieve the toric structure.

The strong paramorphosis between the nematic and the toric textures, that is the persistence in the position of boundaries between domains, could be justified by the presence of local smectic order in the nematic material. Phase transition approaching, the local smectic order experienced by the cybotactic clusters becomes the order of all the material. Smectic plane aggregations in very small regions, as we imagine the cybotactic clusters, can be more persistent near the cell walls and in particular at defects on the wall [7]. We can imagine the defects on the walls to be responsible of the position of the boundaries between domains in the nematic phase. If it is so, it is not surprising an evolution of the nematic texture into smectic as that shown in Fig.8. In this image, we see on the left the cooling of $S_2$ from nematic (139°C) to smectic (138°C) on a range of 1°C. On the right, the corresponding images obtained after digital processing. The image processing was performed with the edge filter of a commercial program. We marked in the figure some regions where there is an evident persistence of boundaries in the phase transition from nematic in smectic.

The results here reported, obtained with the compounds with Br-substituent, confirm the interesting behaviour of the smectic and nematic phases in oxadiazoles. The behaviour is similar to that observed in oxadiazoles with Cl-substituent. In the case of compounds with Br, it is possible to precisely control the nematic-smectic transition, and then we had the opportunity to see the evolution of toric domains. We are planning to use fast image acquisition and verify the behaviour in Cl-substituted compounds.

| Sample | Transition temperatures |
|--------|------------------------|
| $S_1$ | Heating: Cr - 106°C - Sm - 139°C - N - 240°C - I<br>Cooling: I - 238°C - N - 138°C - Sm - 85°C - Cr |
| $S_2$ | Heating: Cr - 102°C - Cr - 110°C - Sm - 123°C - N - 228°C - I<br>Cooling: I - 227°C - N - 120°C - Sm - 80°C - Cr |
| $S_3$ | Heating: Cr - 66°C - N - 93°C - I<br>Cooling: I - 92°C - N - 41°C - Cr |
| $S_4$ | Heating: Cr - 55°C - N - 97°C - I<br>Cooling: I - 96°C - N till room temperature |

**TABLE I: Transition temperatures of the compounds on heating and on cooling.**

**Figure captions**

Figure 1: Molecular structures of the four oxadiazole compounds under investigation.

Figure 2: Toric domains in a cell containing $S_1$ compound. Domains with a large size, approximately 0.05 mm, can be observed (image dimensions 0.38mm x 0.5mm).

Figure 3: Textures observed on cooling compound $S_1$ from the isotropic melt. The image shows on the left the planar texture at a high temperature at 205°C. We see, in the middle, the spherulitic nematic at 145°C, and on the right the toric domains of the smectic phase at 90°C. The texture transition from planar to spherulitic texture in the nematic range is at 150°C.

Figure 4: The texture transition in the nematic melt of compound $S_2$, on heating. The image shows a region with a small spherulitic domain in homeotropic black: at the transition temperature of 175°C, homeotropic texture changes in planar.

Figure 5: The sequence shows how the nematic phase of $S_2$, with spherulitic domains at 121°C, transforms itself in toric smectic texture. The transition is rather slow, occupying in approximately one degree.

Figure 6: Toric domains of smectic $S_2$ in the upper part of the figure. The director field in a vertical cross-section of the cell is drawn in the lower part of the image. The structure of the smectic layers is also drawn. The circle of the focal domain, parallel to cell walls, is represented by the red line; the blue line perpendicular to the cell walls is representing the degenerated hyperbola.

Figure 7: The arrangement of domains is maintained passing from nematic in smectic phase, as shown by the sequence of cooling compound $S_2$, from 121°C (left) to 120°C (right), in the upper part of the image. The molecular anchoring at the walls is preserving the position of boundaries between domains. In the lower part of figure, the drawing sketches the director fields passing from nematic to smectic in two different cases. Case (a) shows the director field in a quasi-planar region surrounded by a homeotropic area, and case (b) the field between two quasi-planar domains. In the domains, the molecules reorient to achieve the toric structure.

Figure 8: Evolution of the nematic texture into smectic for compound $S_1$. On the left the cooling of $S_2$ on a range of 1°C form 139 to 138°C. On the right, the corresponding images obtained with an edge filtering. Some regions where the domain boundaries are persistent in the phase transition are marked in red.

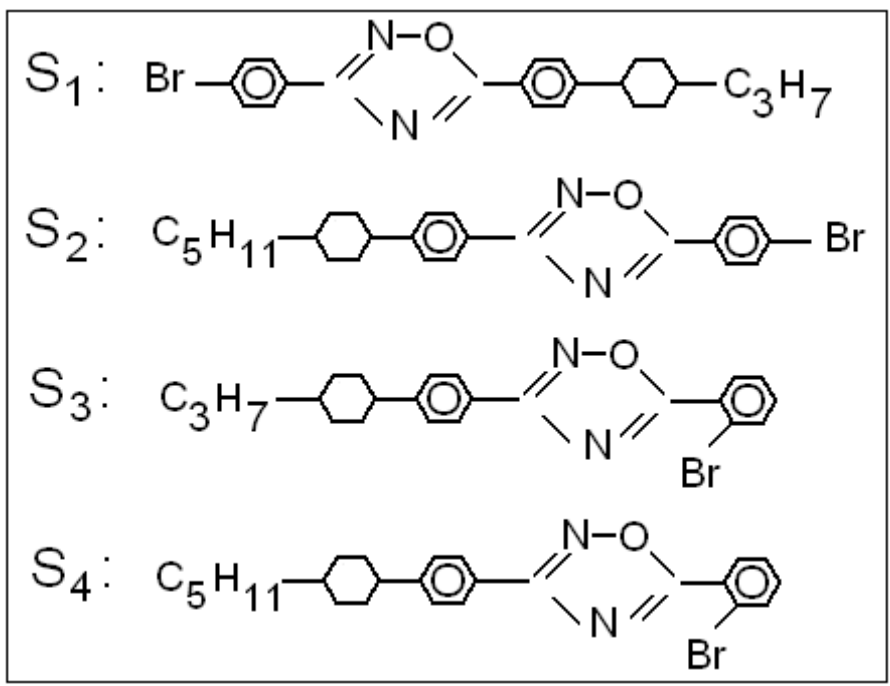

Figure 1

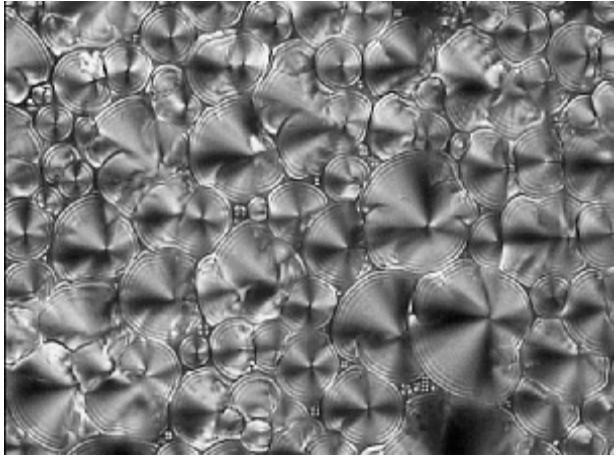

Figure 2

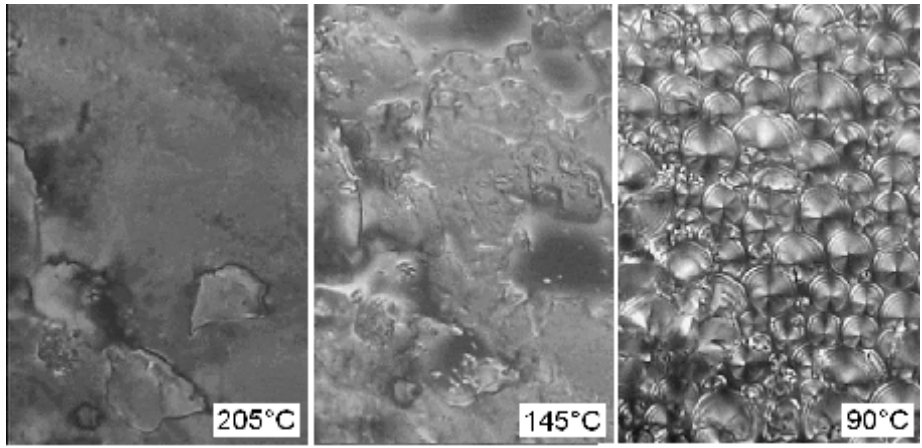

Figure 3

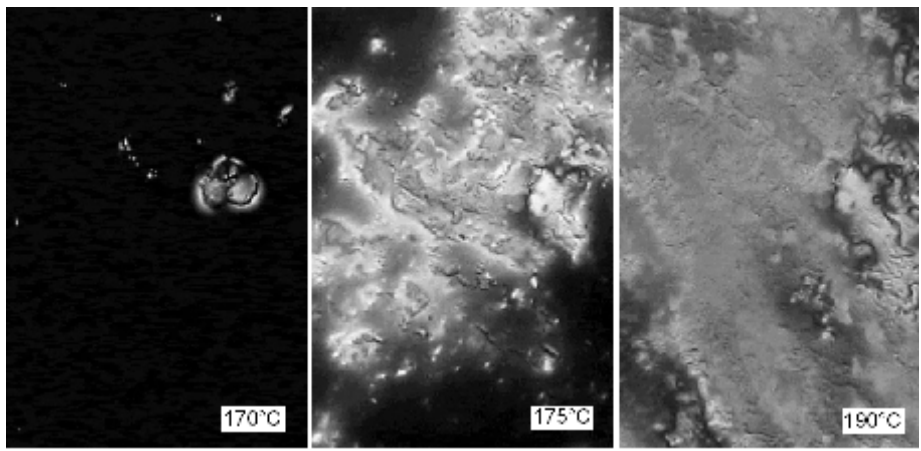

Figure 4

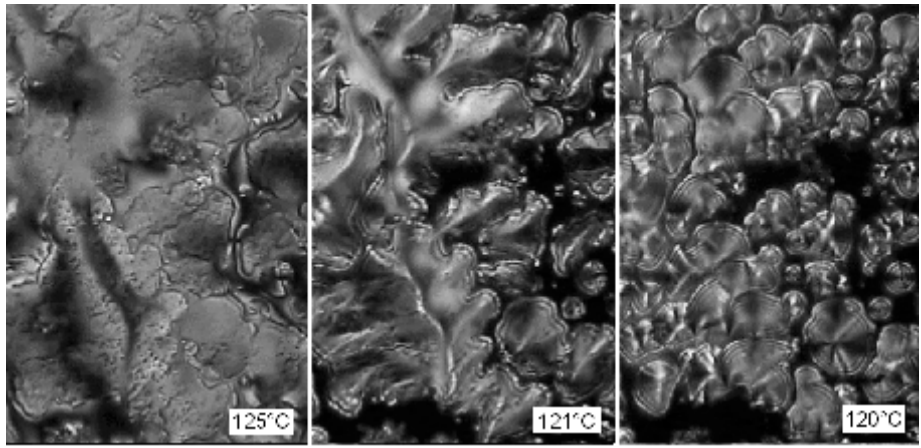

Figure 5

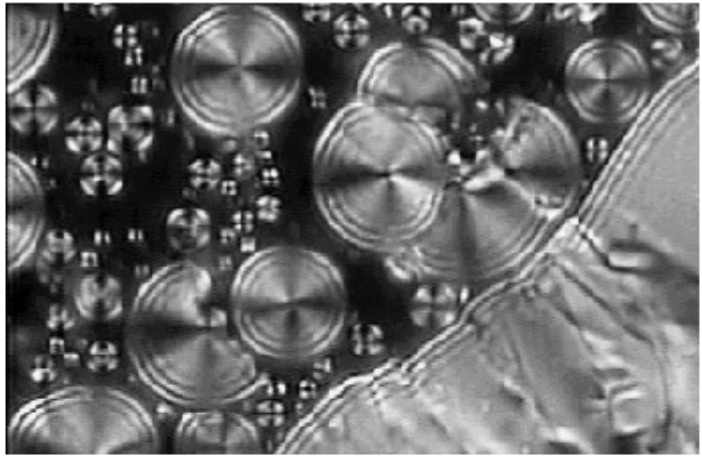

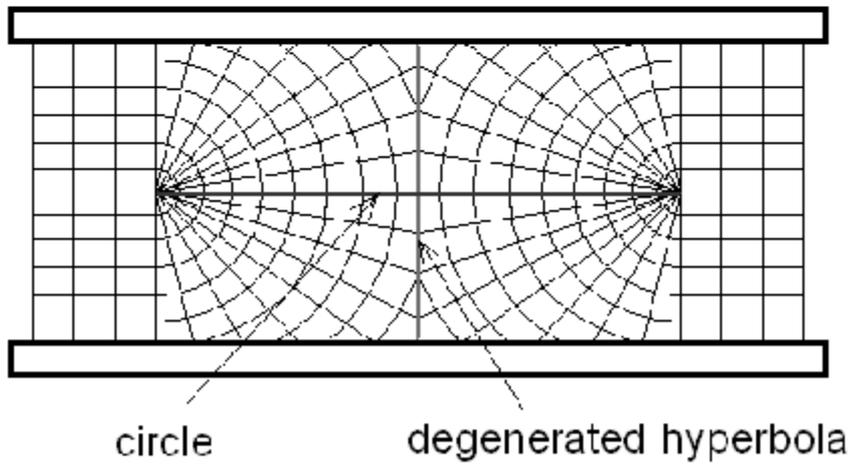

circle    degenerated hyperbola

Figure 6

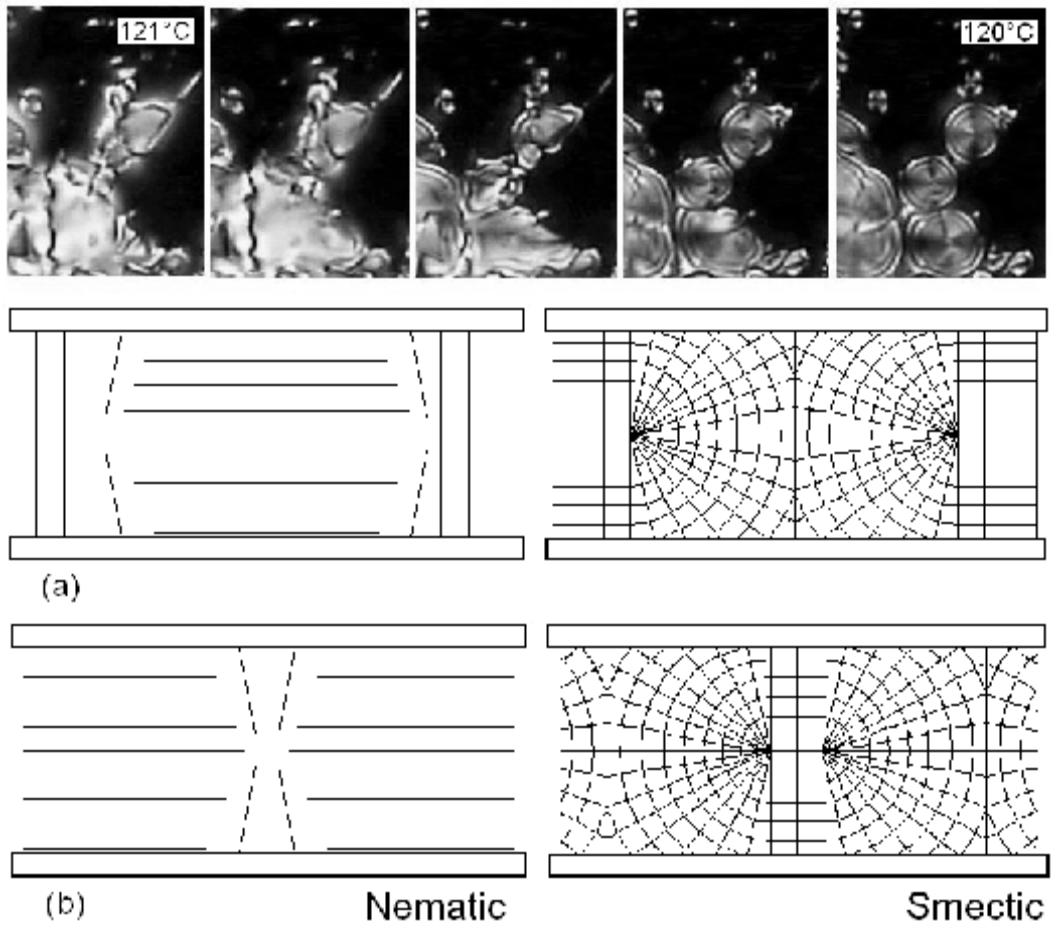

Figure 7

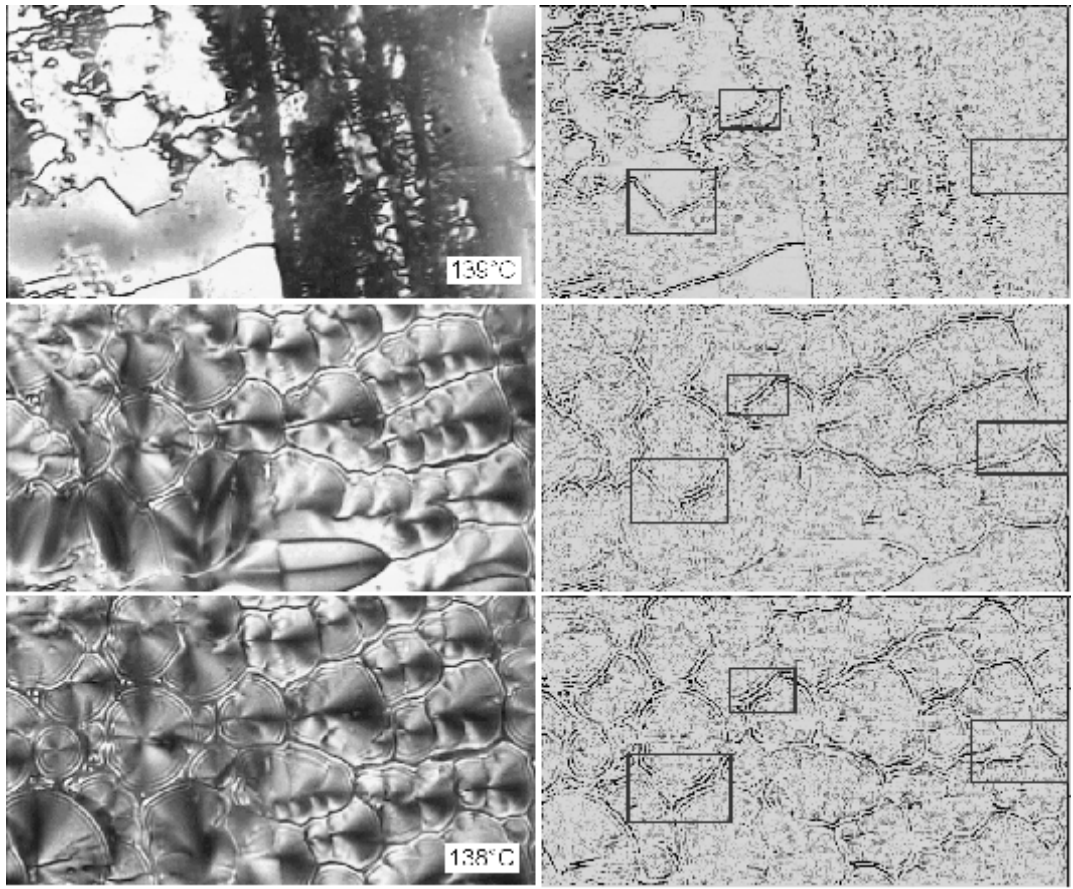

Fig.8